# Geodesics and distance in classical physics.

A. N. Grigorenko

*School of Physics and Astronomy, University of Manchester, M13 9PL, Manchester, UK*

**Abstract:** We formulate geodesics on a manifold in terms of a parallel transfer of a particle state vector transformed by local Lorentz and Yang-Mills symmetry groups. This formulation leads to an introduction of a canonical one-form the eigenvalues of which define distance on a manifold. We suggest an action based on the canonical distance form and apply it to describe classical particles with spin. Arguments are presented in favour of scaling distance in the space-time with a scalar field.

## I. Introduction.

The goal of this paper is to discuss a possible generalization of classical physics that incorporates local symmetries of particle interactions. We achieve this goal by studying geodesics and distance of classical physics. Here we restrict ourselves to classical physics of particles which behaviour is defined by a local state associated with a particle "singularity".

We begin with a brief summary of the paper. In order to describe classical interaction in geometrical terms one needs to introduce additional internal co-ordinates responsible for particle interactions. We assume that these additional co-ordinates are formed by state vectors $|\phi\rangle$ transforming by representations of local Lorentz and Yang-Mills symmetry groups. Hence, we change the space of a classical particle from a co-tangent fibre bundle ($x$, $p$), where $x$ is the particle position and $p$ is the particle momentum, to an associated fibre bundle described by the local co-ordinates ($x$, $|\phi\rangle$), where $x$ are usual space-time co-ordinates and $|\phi\rangle$ are the fibre co-ordinates. Therefore, we reformulate geodesics in terms of the parallel transport of a particle state vector $|\phi\rangle$ (instead of the parallel transfer of a tangent vector). This definition of geodesics leads to an introduction of a one-form $\hat{\theta}$ in an associated fibre bundle and suggests that distance on a manifold, $ds$, can be found as the eigenvalue of this form (described in the Section II and IV) as: $\hat{\theta}(dx)|\phi\rangle = ds|\phi\rangle$. Hence, the quadratic metric form can be replaced by an operator-valued one-form $\hat{\theta}$ that determines the same metrical relations as the metric form and has some analogy with a linear Finsler's metric [1]. The action based on the distance form readily gives a description of classical particles with spin subject to Yang-Mills interactions. The particle state $|\phi\rangle$ replaces the classical particle momentum $p$ in this description. We show that motion of classical spinor particles in our model is affected by the space-time curvature. We also write the distance form of space-time with additional co-ordinates transformed by the local Yang-Mills group of the standard model of electroweak interactions $G$=SU(3)×SU(2)$_L$×U(1)$_Y$ which may be used to describe a classical electron.

The paper is organized as follows. In the Section II we consider a definition of geodesics for particles which states are transformed by local Lorentz and Yang-Mills symmetries. We also introduce a canonical distance form and show how it can be used to define metric relations. In the Section III we provide physical arguments in favour of the canonical distance form. Examples of the canonical distance forms are given in the Section IV. In the Section V we formulate an action principle compatible with the proposed geodesics equations. We show why two definitions of geodesics (based on the parallel transport and the minimal distance) coincide. This action principle allows us to describe classical particles with spin subject to Yang-Mills fields. The distance form of a space-time with the associated fibre of the electroweak theory is proposed in the Section VI and is used to describe a classical electron. Finally, a conclusion is given and the canonical distance form for a space of a Lie group is introduced in the Appendix.

**II. Geodesics definition based on a canonical distance form.**

We start with a reformulation of geodesic equations. Traditionally geodesics are defined through a parallel transport of a tangent vector in a linear (affine) connection on the manifold *M*. (Another definition is based on the fact that a geodesic line is the shortest curve connecting two points on a manifold and will be discussed in Section V.) The geodesic curve $x^i(\tau)$ ($x=\{x^i\}$ is the co-ordinates on a manifold, $i=1,\ldots,n$) can be found from [2, 3]

$$\ddot{x}^i + \omega^i{}_j(\dot{x})\dot{x}^j = 0, \qquad (1)$$

where $\omega^i{}_j$ is the connection forms of the linear connection $\nabla$ (so that the connection form in the principal fibre bundle can be written as $\omega = \omega^i{}_j E_i{}^j$ with $E_i{}^j$ being a basis of the algebra Lie of a linear group [3]), $\dot{x} = dx/d\tau$ is the tangent vector. For natural coordinates $x^i$ we have $\omega^i{}_j = \Gamma^i{}_{jk} dx^k$ where $\Gamma^i{}_{jk}$ are the Christoffel symbols and the geodesics equations in the conventional form are (for natural parameterization)

$$\nabla_{\dot{x}} \dot{x}^i = \ddot{x}^i + \Gamma^i{}_{jk} \dot{x}^j \dot{x}^k = 0 \qquad (2)$$

or simply $\nabla_{\dot{x}} \dot{x} = 0$.

In order to generalize geodesic equation to a space containing particle representations, we consider another definition of geodesics based on horizontal fields [3]. A standard horizontal vector field $B(\xi)$ is defined through a set of equations:

$$\begin{cases} \omega(B(\xi)) = 0 \\ \theta(B(\xi)) = \xi \end{cases},$$

where $\xi$ belong to $\mathbb{R}_n$ ($n$ is the dimension of *M*). Here, $\omega$ and $\theta$ are the connection and canonical forms of the principal linear fibre bundle, respectively [3]. Ref. 3 shows that an integral curve of the horizontal vector field *B* projects onto geodesics under natural projection to *M*. This suggests a definition of geodesics as a solution of a system

$$\begin{cases} \nabla_{\dot{x}} u = \dot{u} + \hat{\omega}(\dot{x}) \cdot u = 0 \\ \theta(\dot{x}) = u \end{cases}, \qquad (3)$$

where *u* is a $\mathbb{R}_n$-vector generated by the canonical form $\theta(\dot{x})$ and $\nabla_{\dot{x}} u$, $\omega(\dot{x})$, $\theta(\dot{x})$ are the interior products $\iota_{\dot{x}} \nabla u$, $\iota_{\dot{x}} \omega$, $\iota_{\dot{x}} \theta$. The linear canonical form of a manifold is given by $\theta = \theta^i e_i$, where $e_i$ are basis vectors of the tangent space and $\theta^i$ are the canonical one-forms of the linear connection [3]. For natural coordinates $x^i$, the canonical form of linear connection is $\theta^i = dx^i$ and $e_i = \dfrac{\partial}{\partial x^i}$. Then, the second equation of (3) is $u^i = \dot{x}^i$. Hence we can eliminate the tangent vector *u* and write the geodesics equations in the conventional form (2). It is worth noting that the system (3) has a transparent physical meaning. The first equation of (3) implies that a tangent vector *u* is transported parallel along a geodesic curve. The second equation of the system (3) determines a displacement on the manifold for a specific value of a tangent vector, or, in other words, solders the manifold with the tangent space.

Equations (1)-(3) determine geodetic motion for a structureless material point. However, there exists strong experimental evidence that physical objects are not just material points. Instead, they possess internal degrees of freedom transforming under representations of the Lorentz group and a Yang-Mills group of local symmetry. (We can safely say that all our knowledge of the universe comes from fibre bundles associated with particles and their states.) Therefore, it is necessary to reformulate geodesic equations in terms of representations of these groups. From this point onwards we suppose that the configuration space of classical physics is an associated fibre bundle with local co-ordinates $(x, |\phi\rangle)$, where *x* are the space-time co-ordinates and $|\phi\rangle$ are the fibre co-ordinates (state vectors transforming by representations of local Lorentz and Yang-Mills symmetries).

First, we rewrite geodesic equations in terms of representations $|\phi\rangle$ of the Lorentz group. To perform this task we have to reduce the linear connection of arbitrary frames to the linear connection of the orthonormal frame rotations since only orthonormal frames support the action of the Lorentz group. Let $e_a$ be a section of orthogonal frames so that $e_i = e_a e^a{}_i$, where $e^a{}_i$ are *n*-ads (*n* is the space dimension), and $\omega^a{}_b$ be the connection forms $\omega^i{}_j$ reduced to the orthonormal frames $e_a$: $\omega^a{}_b = e^a{}_j \omega^j{}_i e_b{}^i + e^a{}_i de_b{}^i$, where $e_b{}^i$ is the inverse to *n*-ads. We describes the particle state using a representation vector $|\phi\rangle$ assigned to the particle position so that $|\phi(\tau)\rangle$, where $\tau$ is the curve parameter. (In the conventional definition of geodesics $|\phi\rangle$ is the tangent vector $u$ assigned to the particle position on the manifold). Then, the first equation of the system (3) has a straightforward generalisation to a representation of the Lorentz group

$$\nabla_{\dot{x}} |\phi\rangle = d|\phi\rangle / d\tau + \hat{\omega}(\dot{x}) |\phi\rangle = 0. \tag{4}$$

Here $|\phi\rangle$ is a vector of the representation, $\hat{\omega} = (1/2)\hat{G}_{[ab]} \omega^{ab}$ is the linear connection form in the representation, $\hat{G}_{[ab]}$ are the generators of the Lorentz group in the same representation [4]. Equation (4) simply tells that any vector object associated with a geodesic curve should be transported parallel along it and is valid for representations of a local Yang-Mills group as well.

The difficulty lies with the second equation of the geodesic system (3) (that calculates the direction and magnitude of the displacement on a manifold for a specific value of the tangent vector and solders manifold with its tangent space). This equation cannot be formulated in terms of representation vectors since one-to-one correspondence between representation vectors $|\phi\rangle$ and displacements on a manifold is absent in a general case. It has to be said that the second equation of the geodesic system (3) is intuitively obvious to such degree that it is usually taken for granted. However, it is not necessarily evident. Generally speaking, it relates two entities of *different geometrical nature*: one is connected to motion along a curve on a manifold $x(\tau)$ (a non-local property) and the other is a tangent vector $u$ (a local property). There is a long-standing discussion among mathematicians (initiated by Newton and Leibniz) about a possibility to identify the tangent space to a manifold with the manifold itself [3]. There are two basic views on this subject. The first approach (loosely attributed to Newton) is that small displacements *are* the tangent vectors taken at the point $x(\tau)$, the canonical form $\theta$ is the identity on the tangent space and the geodesics are given by (1), (3). This approach is obviously correct. However, it is difficult to see how one can generalise this approach and take into account other possible degrees of freedom of physical objects. The second approach (loosely attributed to Leibniz) suggests that the tangent space is a space of derivations [3] and is not automatically related to displacements on the underlying manifold. In this approach the second equation of the system (3) is just one possibility allowing one to solder manifold with its tangent space. Physicists deal with the same problem formulated in different terms. A definition of an instant velocity (a tangent vector) based on space-time displacements is elusive [5]. It is not clear what time interval one needs to watch an object in order to deduce its instant velocity from motion in space-time nor what is the point of space-time which this instant velocity belongs to. For these reasons, we believe that the second equation of (3) has to be amended in order to describe geodesics in terms of representations of local group of symmetries.

To find a correct relation between displacements on a manifold and representation vectors we note that representation operators are the only geometric objects at our disposal. A fundamental property of a (physical) operator is a set of its eigenvectors lying in the representation space. Thus, an operator valued one-form (which maps a displacement to a space of operators) would establish a mapping of a manifold displacement to eigenvectors of a corresponding operator. This mapping is multivalued since different eigenvectors may correspond to the same displacement. However, being taken with equation (4), this correspondence may lead to a unique curve in the fibre bundle associated with the representation and yield a geodesic as a projection of this curve.

For concreteness, we consider the lowest spinor representation $|\psi\rangle$ of the Lorentz group (the generalization to higher representations is straightforward). It is known that a spinor representation possesses a set of $\hat{\gamma}_a$ matrices with property

$$\hat{\gamma}_a\hat{\gamma}_b + \hat{\gamma}_b\hat{\gamma}_a = 2\eta_{ab}I_n, \qquad (5)$$

where $\eta_{ab}$ is the metric in an orthogonal frame, $I_n$ is the identity matrix and $n$ is the dimension of the manifold. These matrices generate a spinor representation $S(\Lambda)$ by

$$\Lambda_a{}^b\hat{\gamma}_b = S^{-1}(\Lambda)\hat{\gamma}_a S(\Lambda), \qquad (6)$$

where $\Lambda = \|\Lambda_a{}^b\|$ is a matrix of an Lorentz transformation and $S(\Lambda)$ is the corresponding operator in the spinor space of $|\psi\rangle$. We suggest that an operator-valued one-form

$$\hat{\theta} = \hat{\gamma}_a \theta^a, \tag{7}$$

where $\theta^a$ are the displacement one-forms $\theta^i$ of the canonical form $\theta$ of the linear connection taken with respect to orthonormal frames $e_a$ (so that $\theta = \theta^a e_a$), indeed provides a desired relation between displacements and the representation vectors. Then, the geodesics curve $x^i(\tau)$ in terms of the particle representation vectors can be written as a solution of the system

$$\begin{cases} \nabla_{\dot{x}} |\psi\rangle = 0 \\ \hat{\theta}(\dot{x}) |\psi\rangle = \dot{\lambda} |\psi\rangle \end{cases}, \tag{8}$$

where $\lambda(\tau)$ is a scalar parameter and $\nabla_{\dot{x}}|\psi\rangle$, $\hat{\theta}(\dot{x})$ stands for $\iota_{\dot{x}}\nabla|\psi\rangle$, $\iota_{\dot{x}}\hat{\theta}$ respectively, and $\dot{\lambda}$ is an eigenvalue of the form $\hat{\theta}(\dot{x})$. The first equation of the system (8) tells that the representation vector $|\psi\rangle$ is transported parallel along the curve. The second equation of the system (8) implies that the representation vector $|\psi\rangle$ stays an eigenvector of the form (7) while being transported parallel. If the form $\hat{\theta}(\dot{x})$ has several eigenvalues, the maximal one $\dot{\lambda}_{max}$ has to be chosen.

We prove now that a solution of (8) ($|\psi(\tau)\rangle$, $x(\tau)$) indeed projects onto a geodesic provided the linear connection $\theta^a$, $\omega^a{}_b$ ($\theta^i$, $\omega^i{}_j$) is torsion free. (This proof is just a simple generalization of the derivation of geodesics from integral curves of horizontal fields [3].) The first equation of (8) can be integrated for any curve $x(\tau)$ on the manifold and lifts the curve into the fibre bundle associated with the representation space. Taking the covariant derivative of the second equation of (8) and using the first one, we find that system (8) is compatible if $\nabla_{\dot{x}}\hat{\theta}(\dot{x})|\psi\rangle = 0$ (in the natural calibration where $\dot{\lambda}_{max} = 1$). It is easy to see that the condition

$$\nabla_{\dot{x}}\hat{\theta}(\dot{x}) = 0 \tag{9}$$

is sufficient for (8) to be compatible and to have a solution. Indeed, (9) means that the operator $\hat{\theta}$ is transported parallel along the solution curve $x(\tau)$ and the relation $\hat{\theta}(\dot{x})|\psi\rangle = \dot{\lambda}_{max}|\psi\rangle$ is conserved for the representation vector $|\psi\rangle$ that is transported parallel according to the first equation of (8). Finally, writing $\hat{\theta}$ as $\hat{\theta}(\dot{x}) = \hat{\gamma}_a e^a{}_i \dot{x}^i$ and noting that the covariant derivatives of $\hat{\gamma}$ matrices and n-ads $e^a{}_i$ vanish for the spinor connection induced by any torsion free linear connection [4], $\nabla \hat{\gamma}_a = 0$, $\nabla e^a_{\ i} = 0$, we see that (9) is equivalent to the condition $\nabla_{\dot{x}} \dot{x}^i = 0$, which is the geodesic equation of the manifold in a natural parameterization. It means that a geodesic $\nabla_{\dot{x}} \dot{x}^i = 0$ and the parallel transport of $|\psi\rangle$ along the geodesic is the solution of the system (8). Vice versa, if the system (8) has a solution it satisfies (9) and hence projects onto a geodesic.

Since operator-valued one-form $\hat{\theta}$ is transformed by the adjoint representation of the Lorentz group (in accordance with the adjoint representation of $\hat{\gamma}_a$), we can express the torsion-free condition for $\hat{\theta}$ as

$$\nabla \hat{\theta} = d\hat{\theta} + [\hat{\omega}, \hat{\theta}] = 0. \qquad (10)$$

Direct calculations show that (10) is valid for any torsion-free linear connection ($\theta = \theta^a e_a$, $\omega = \omega^a_{\ b} E_a^{\ b}$ and $\nabla \theta^a = d\theta^a + \omega^a_{\ b} \wedge \theta^b = 0$). It means that we can regard $\hat{\theta}$ as the canonical form of the linear connection in spinor representation. Also, the relations (7) and (5) imply that

$$\mathbf{g} I_n = \hat{\theta} \cdot \hat{\theta}, \qquad (11)$$

where $\mathbf{g}$ is the metric two-form. Applying the operator $\hat{\theta}$ to the second equation of (8) and using (11), we see that the eigenvalues $d\lambda = \pm ds$ of the form $\hat{\theta}$ are given by the metric generated distance $ds = \sqrt{\eta_{ab} \theta^a \theta^b} = \sqrt{g_{ik} dx^i dx^k}$. For these reasons the operator-valued one-form $\hat{\theta}$ will be referred to as the distance form (or the canonical distance form). It is clear from (11) that the distance form is a natural way to extract the square root from the metric two-form. In the coordinate basis the form $\hat{\theta}$ is written as $\hat{\theta} = \hat{\gamma}_i dx^i$, where the matrices $\hat{\gamma}_i = \hat{\gamma}_a e^a_{\ i}$ satisfy the following relations: $\hat{\gamma}_i \hat{\gamma}_j + \hat{\gamma}_j \hat{\gamma}_i = 2 g_{ij} I_n$ and $\nabla \hat{\gamma}_i = 0$.

It is easy to check that the distance form is enough to recover a metric tensor, a torsion–free linear connection compatible with the metric and the curvature. Indeed, the operator-valued form $\hat{\theta}$ immediately yields the metric with the help of (11), $\mathbf{g} = (1/n) \mathrm{Tr}(\hat{\theta} \cdot \hat{\theta})$. The displacement one-forms $\theta^a$ are given by $\theta^a = (1/n) \mathrm{Tr}(\hat{\gamma}^a \hat{\theta})$. The reciprocal basis of frames $e_a$ can be calculated from the relations $\langle \theta^a, e_b \rangle = \delta^a_{\ b}$, where $\langle , \rangle$ denote the scalar product of one-forms with vectors ($\delta^a_b$ is the Kronecker symbol). Next, we introduce the operator-valued basis of frames $\hat{e} = \hat{\gamma}^a e_a$ that is transformed by the adjoint representation of the orthogonal group $\hat{e}' = \hat{\gamma}^a \Lambda_a{}^b e_b = S^{-1}(\Lambda)\hat{\gamma}^a e_a S(\Lambda) = S^{-1}(\Lambda)\hat{e}S(\Lambda)$ and contains all information about frames $e_a = (1/n)\mathrm{Tr}(\hat{\gamma}_a \hat{e})$. The covariant derivative of the operator-valued basis is zero $\nabla \hat{e} = d\hat{e} + [\hat{\omega}, \hat{e}] = 0$. We also introduce the operator-valued (non-associative) scalar product between the operator-valued forms $\hat{f} = \hat{p}_a \theta^a$ and vectors $\hat{v} = \hat{q}^a e_a$ by the relation $\langle \hat{f}, \hat{v} \rangle = \hat{p}_a \hat{q}^b \langle \theta^a, e_b \rangle = \hat{p}_a \hat{q}^a$. Differentiating the identity $\langle \hat{\theta}, \hat{e} \rangle = nI_n$ and using the fact that the covariant derivatives of the canonical metric form and the operator-valued basis vanish, we get the connection $\hat{\omega}$ as

$$\hat{\omega} = \frac{1}{4} \langle [d\hat{\theta}, \hat{e}] \rangle, \tag{12}$$

where [,] stands for commutation. It allows us to recover the connection form $\omega_{ab} = (2/n)\mathrm{Tr}(\hat{G}_{[ab]} \hat{\omega})$ with $\hat{G}_{[ab]} = (1/4)(\hat{\gamma}_a \hat{\gamma}_b - \hat{\gamma}_b \hat{\gamma}_a)$ and yields the curvature in the spinor representation as

$$\hat{R} = \frac{1}{16} \Big( \langle d\hat{\theta}, [\hat{e}, \langle [d\hat{\theta}, \hat{e}] \rangle ] \rangle + [\langle [d\hat{\theta}, \hat{e}] \rangle, \langle [d\hat{\theta}, \hat{e}] \rangle ] \Big). \tag{13}$$

The formulae (12), (13) are not surprise since the metric form **g** also defines these geometric properties. They mean, however, that the connection distance form $\hat{\theta}$ is enough to recover basic properties of the linear connection on a manifold.

Using Infeld-van der Waerden symbols [6], we can construct the distance form for any representation produced by the direct product of spinor representations (provided the product of these representations with the vector representation contains the trivial representation), e.g., for massive particles described by an equation $\left( \hat{\Gamma}^a \frac{\partial}{\partial x^a} \right) |\phi\rangle = m |\phi\rangle$, where $\Gamma^a \frac{\partial}{\partial x^a}$ is transformed by the adjoint representation of the Lorentz group. In this case eigenvalues of the operator-valued one-form $\hat{\Gamma}_a dx^a$ are conserved under Lorentz transformations and should be proportional to distance. Hence, the system (8) with $\hat{\gamma}_a = \hat{\Gamma}_a$ and an appropriate connection form $\hat{\omega}$ will give geodesics.

To summarize, the definition of geodesics in a fibre bundle of a Lorentz (and generally speaking orthogonal) group associated with $|\phi\rangle$-representation can be given as follows: *geodesic is a curve such that the parallel transport of the initial representation vector $|\phi\rangle$ to any point along the curve yields an eigenvector of the operator-valued distance form $\hat{\theta}(\dot{x})$ taken at this point* (in accordance with the original definition where a geodesics is defined as the curve such that the parallel transport of a tangent vector gives a displacement on the manifold via canonical forms $\theta^i$). Since both $\hat{\theta}$ and the covariant derivative $\nabla$ are transformed by the adjoint representation of the orthogonal group, the definition (8) does not depend upon the calibration (a local section of the fibre bundle of the orthogonal group associated with the representation). The eigenvalues of the distance form coincide with the metric generated distance. This will allow us to formulate geodesics as curves of minimal lengths, see the Section V.

It is clear that the geodesic definition (8) is obviously valid when $|\phi\rangle$ is also transformed by a Yang-Mills group commuting with the local orthogonal group.

**III. Physical arguments for a distance form.**

In this section we present physical arguments for the case that distance should be regarded as an eigenvalue of a suitably chosen distance one-form operator. First, we make a trivial observation that empty space does not possess any metrical relations relevant to physics. It is evident that some phenomena are required to mark points of space-time and deduce distances between them. Thus, being a good hypothetic background to the Newton's theory, a concept of empty space-time cannot be applied to the real world. It is widely accepted that our world possesses coordinates additional to space and time and these coordinates describe physical phenomena that happens in the world. At the present state of knowledge we associate some properties of these coordinates with particles and fields. We assume that every point of our space-time has its own copy of additional coordinates, which allows us to treat the world as a fibre bundle [7, 8]. The base of this fibre bundle is a four-dimensional space-time which is an arena of the low-energy physics. The fibres are the additional coordinates that can be imagined as representations of a local symmetry group. It is supposed that locally physics is simple and the fibre space from one point of space-time can be connected with the fibre of an adjacent point by means of a linear connection. System dynamics does not depend upon the way in which additional coordinates are parameterised, which finds its reflection in gauge invariance of all measured quantities. There are different opinions to what are particles and what are fields in this picture. In classical physics fields are treated as local objects characterised by their potentials (local functions of space-time coordinates). The field potentials are associated with connections in the world fibre bundle and fields are related to the curvatures of these connections [7, 8]. Particles are supposed to be non-trivial states of the world fibre bundle, which are described by globally non-trivial connections [7, 8]. Since any non-trivial state of the world fibre bundle is accompanied by non-trivial connections, it means that a (classical) particle is surrounded by fields associated with it. An electron, for example, is surrounded by an electromagnetic field. This implies that particles, generally speaking, are non-local objects.

Many scientists realised that the metric properties of space-time should be deduced from properties of constituent matter. However, it required Einstein's genius to show how physical phenomena can be used to measure point separations in space-time [9]. Einstein utilised light signals (photons) for space-time mapping and measuring distances. His light signal procedure works very well for sufficiently large distances. However, measurement of small separations with the help of photons is hindered by the finite wavelength of the light and by the problem of the photon creation and absorption in small space-time regions. It is very unlikely that a very small photon emitter and receiver can be fabricated. This means that small distances cannot be measured with the help of light signals and hence there is a necessity to revise the procedure of distance measurement.

The only alternatives to massless photons (for mapping and measuring space) are massive particles. Despite we do not know an internal structure of elementary particles, we believe that a massive particle has some sort of singularity which is well localised in space. This singularity may be used to probe space-time with better resolution than that achieved with light signals. Again, we can dismiss any direct distance measurement procedure associated with the creation of a particle at some point of space-time and its annihilation at some other point for the same difficulty of fabrication of very small particle emitters and absorbers. As a result, for probing space-time with high resolution, we are forced to use indirect procedures based on a change of a state of an already existing or specially prepared particle. Indeed, information about a particle state and position can be obtained by recording particle fields with the help of a macroscopic apparatus placed sufficiently far away from the region where distance is being assessed. For example, in classical physics we can determine a position of an electron by measuring its electromagnetic field in several different points of space-time sufficiently far from the electron location.

Therefore, it seems natural to consider the following procedure for distance determination: we prepare a physical particle that travels from a given point of space-time to a neighbouring point and deduce the distance between these two close points by measuring the fields produced by the particle with the help of a macroscopic apparatus. We need only to specify what we imply by "particle preparation". It is evident that once we refuse to consider particles as structureless material points we allow a particle to transform during its motion. There are two basic types of transformations. One type is passive transformations which come from an agreement on "what is what" in a fibre space over a fixed point of the manifold. In mathematical terms this agreement means freedom to choose any point of a fibre as the beginning of additional coordinates, or, in other words, freedom of gauge fixing. Another type of transformations consists of active transformations which relate to a real change of physical particles. An electron, e.g., can emit photons, photons can generate electron-positron pairs, etc. So, we formulate the basic (identity) principle of the particle preparation for distance measurement: *a particle should stay in the same state during the act of measurement*. This identity principle ensures that we observe the same test particle during an entire act of measurement and hence possible "stray" particles do not influence our results. Also, this principle guarantees that the change of the world fibre bundle (caused by the presence of the particle) will be the same in the beginning and in the end of measurement. (The presence of a physical particle influences the state of the world fibre bundle. The identity principle ensures that this influence is homogeneous.)

In mathematical terms the identity principle for distance measurement with respect to passive transformations means that the state of the particle should be conserved in process of measurement, so that observers in different points of space-time can agree that they "see" the same particle from the point of view of gauge fixing:

$$\nabla |\phi\rangle = 0, \qquad (14)$$

where $|\phi\rangle$ is the particle state and $\nabla$ is the connection for this particle state. With respect to active particle transformations the identity principle implies that a particle state should be an eigenstate of some distance operator $\hat{L}(dx^i)$ (distance measurement should not change a physical state of the particle):

$$\hat{L}(dx^i)|\phi\rangle = d\lambda|\phi\rangle. \qquad (15)$$

This is as far as we can go without describing the particles state $|\phi\rangle$. In general, particles (combined with accompanying fields) are non-local objects. Yet, for distance measurement, we need to associate a particle state to some point in space-time. A way to do that in classical physics is to attribute a representation vector to the place of particle singularity (where the fields are "diverging"). The main argument in favour of this is based on the fact that the measured distance should not depend on the internal structure of the probing particle and should not change if we "simplify" the structure of the particle to a local representation vector. Then, $|\phi\rangle$ will be a vector of a representation of an orthogonal group and the equation (14) will correspond to the first equation of the geodetic system (8). For the distance operator $\hat{L}(dx^i)$ we argue that the linear operator (7) is the simplest one that gives distance consistent with the Riemann metric.

In conclusion to this section, we stress again that metrical relations (in the physical world) should be deduced from properties of constituent matter. Since particles and fields are the only ingredients of the contemporary physics and it is highly unlikely that fields can be used for measuring small separations, distance should be directly related to some properties of physical particles. This forces us to suggest that distance is given by eigenvalues of some distance one-form instead of the conventional two-form acting on the tangent space.

The fact that distance is an eigenvalue of a distance operator does not necessarily mean quantisation of distance and volume. In this classical picture it rather implies that we have to prepare the measuring apparatus (elementary particles used for distance measurement) before making a conclusion about separation of two points in space-time. The quantum mechanics will probably require more complex particle states and complex distance operators.

**IV. Examples of the canonical distance form.**

***Three-dimensional Euclid space, n = 3, metric signature (+,+,+).***

The two-component spinors realise the lowest spinor representation in three-dimensions. The canonical distance form in the spinor representation is

$$\hat{\theta} = \hat{\sigma}_a \theta^a \tag{16}$$

where $\hat{\sigma}_a$ ($a = 1,2,3$) are the Pauli matrices. They satisfy the well-known relation

$$\hat{\sigma}_a \hat{\sigma}_b + \hat{\sigma}_b \hat{\sigma}_a = 2\delta_{ab} \tag{17}$$

where $\delta_{ab}$ is the Euclid metric. The canonical distance form (16) has eigenvalues of $\pm dr = \sqrt{\delta_{ab} \theta^a \theta^b}$. In flat Euclid space the canonical distance form is $\hat{\theta} = \begin{pmatrix} dz & dx - idy \\ dx + idy & -dz \end{pmatrix}$. In three dimensions there exists a one-to-one correspondence between direction of motion on the manifold and the canonical form eigenvector (up to a phase factor). Indeed, if $\iota_{\delta x} \theta^a = \delta x^a$ and $(\delta x^1, \delta x^2, \delta x^3) = dr(\sin\theta\cos\varphi, \sin\theta\sin\varphi, \cos\theta)$, where $\varphi, \theta$ are the polar and azimuthal angles, then, the eigenvector which corresponds to this displacement is

$$|\chi_{\delta x}\rangle = \begin{pmatrix} \cos(\theta/2)\exp(-\iota\varphi) \\ \sin(\theta/2) \end{pmatrix}. \tag{18}$$

This correspondence is directly connected with the possibility to pack the distance form into the Lie algebra of the orthogonal group in three dimensions. In general case, the canonical distance form and the generators of the Lie algebra of an orthogonal group lie in different sub-spaces of the Clifford algebra: one sub-space is generated by $\hat{\gamma}_a$ matrices while the other is generated by the commutators $\hat{G}_{[ab]} = (1/4)(\hat{\gamma}_a \hat{\gamma}_b - \hat{\gamma}_b \hat{\gamma}_a)$. However, in three dimensions, the Levi-Civita totally antisymmetric tensor connects these two subspaces. As a result, matrices $\hat{\sigma}^a = -(i/2)\varepsilon^{abc}(\hat{\sigma}_b \hat{\sigma}_c - \hat{\sigma}_c \hat{\sigma}_b)$ can be represented as linear combinations of generators of the Lie group in the spinor representation. Analogously, the canonical distance form can be moved into the Lie algebra of the orthogonal group as $\hat{\theta} = i \cdot ad(\hat{X}_a)\theta^a$, where $\hat{X}_a$ are viewed as operators acting on the Lie algebra by the adjoint representation: $ad(\hat{X}_a)\hat{X}_b = [\hat{X}_a, \hat{X}_b] = c^d{}_{ab}\hat{X}_d$ (here $c^d{}_{ab}$ are the structural constants of the group). It is easy to check that the form $\hat{\theta} = i \cdot ad(\hat{X}_a)\theta^a = i\hat{c}_a \theta^a$, where $\hat{c}_a = \|c^d{}_{ab}\|$, indeed has eigenvalues of $\pm dr = \sqrt{\delta_{ab}\theta^a \theta^b}$. In three dimensions, the generators $\hat{X}_a$ realise the same vector representation $D_1$ of the orthogonal group as the tangent vectors. Yet, the canonical distance form $\hat{\theta} = i \cdot ad(\hat{X}_a)\theta^a = i\hat{c}_a\theta^a$ is not an "identity" ($\hat{\theta} \neq \hat{X}_a\theta^a$) in the vector space of generators (in contrast to the space of tangent vectors where $\theta = e_a\theta^a$).

## Four-dimensional Lorentz space-time, n = 4, metric signature (+,-,-,-).

The Dirac bi-spinors realise the lowest spinor representation in the Lorentz space-time and the distance form is written as

$$\hat{\theta} = \hat{\gamma}_a\theta^a, \qquad (19)$$

where $\hat{\gamma}_a$ ($a = 0,1,2,3$) are Dirac matrices. (In flat space the canonical distance form is

$$\hat{\theta} = \begin{pmatrix} 0 & 0 & dt+dz & dx-idy \\ 0 & 0 & dx+idy & dt-dz \\ dt-dz & -dx+idy & 0 & 0 \\ -dx-idy & dt+dz & 0 & 0 \end{pmatrix}).$$

In four-dimensions the spinor representation is a direct sum of two irreducible representations of the orthogonal group $|\psi\rangle = \chi \oplus \eta$, where $\chi$ is transformed under (1/2,0) and $\eta$ is transformed by (0,1/2) representations of the orthogonal group [6]. We shall use the following notations throughout the paper: $L = \chi$, $R = \eta$ and $|\psi\rangle = L \oplus R = \begin{pmatrix} L \\ R \end{pmatrix}$, where $L$ and $R$ and the left and right components of the Dirac bi-spinor. Since $|\psi\rangle$ is the direct sum of two irreducible components, the distance form (the canonical form in the spinor representation) is also a direct sum of two operator-valued one-forms connecting spaces of left and right chirality

$$\hat{\theta} = \hat{\theta}_{RL} \oplus \hat{\theta}_{LR} = \begin{pmatrix} 0 & \hat{\theta}_{LR} \\ \hat{\theta}_{RL} & 0 \end{pmatrix}, \qquad (20)$$

where $\hat{\theta}_{RL} = \hat{\sigma}_a\theta^a$ and $\hat{\theta}_{LR} = \hat{\tilde{\sigma}}_a\theta^a$ ($a = 0,1,2,3$) with $\hat{\sigma}_a = (I_2, \hat{\sigma}_m)$, $\hat{\tilde{\sigma}}_a = (I_2, -\hat{\sigma}_m)$, $I_2$ is the two-dimensional identity matrix and $\hat{\sigma}_m$ ($m = 1,2,3$) are the Pauli matrices. As a consequence of this decomposition the square of the distance form is

$$\hat{\theta} \cdot \hat{\theta} = \hat{\theta}_{RL} \cdot \hat{\theta}_{LR} \oplus \hat{\theta}_{LR} \cdot \hat{\theta}_{RL} = \begin{pmatrix} \hat{\theta}_{LR} \cdot \hat{\theta}_{RL} & 0 \\ 0 & \hat{\theta}_{LR} \cdot \hat{\theta}_{RL} \end{pmatrix} = ds^2\begin{pmatrix} I_2 & 0 \\ 0 & I_2 \end{pmatrix}, \text{ where}$$

$$\hat{\theta}_{RL} \cdot \hat{\theta}_{LR} = \hat{\theta}_{LR} \cdot \hat{\theta}_{RL} = \hat{\sigma}_a\theta^a \cdot \hat{\tilde{\sigma}}_b\theta^b = \hat{\tilde{\sigma}}_a\theta^a \cdot \hat{\sigma}_b\theta^b = \eta_{ab}\theta^a\theta^b I_2 = ds^2 I_2. \qquad (21)$$

The connection form $\hat{\omega}$ is $\hat{\omega} = \hat{\omega}_{LL} \oplus \hat{\omega}_{RR} = \begin{pmatrix} \hat{\omega}_{LL} & 0 \\ 0 & \hat{\omega}_{RR} \end{pmatrix}$, where $\hat{\omega}_{LL} = (1/4)\hat{\tau}_{ab}\omega^{ab}$, $\hat{\omega}_{RR} = \hat{\tilde{\tau}}_{ab}\omega^{ab}$ with $\hat{\tau}_{0m} = -\hat{\tau}_{m0} = \hat{\sigma}_m$, $\hat{\tilde{\tau}}_{0m} = -\hat{\tilde{\tau}}_{m0} = -\hat{\sigma}_m$, $\hat{\tau}_{pq} = \hat{\tilde{\tau}}_{qp} = -i\varepsilon_{pqm}\hat{\sigma}_m$ ($m, p, q = 1,2,3$). The condition of the absence of torsion is $d\hat{\theta}_{LR} = \hat{\omega}_{LL} \wedge \hat{\theta}_{LR} - \hat{\theta}_{LR} \wedge \hat{\omega}_{RR}$ and $d\hat{\theta}_{RL} = \hat{\omega}_{RR} \wedge \hat{\theta}_{RL} - \hat{\theta}_{RL} \wedge \hat{\omega}_{LL}$.

Writing the distance form as $\hat{\theta} = ds(\hat{Q}^2 \oplus \hat{Q}^{-2})$ with $\hat{Q} = \sqrt{\hat{\theta}_{RL}/ds}$ (which is possible due to (21)), we find that the eigenvalues of the distance form (19) are indeed intervals $\pm ds$ with the eigenvectors being

$$|\psi_\pm\rangle = \frac{1}{\sqrt{2}}\left(Q^{-1}\chi \oplus \pm Q\chi\right) = \frac{1}{\sqrt{2}}\begin{pmatrix} Q^{-1}\chi \\ \pm Q\chi \end{pmatrix}, \qquad (22)$$

where $\chi$ is any normalised spinor. The freedom to choose any state of $\chi$ while determining the distance reflects the fact that (spinor) particle representations in 4 dimensions possess spin degrees of freedom.

In this case the canonical distance form cannot be packed into the Lie algebra of the orthogonal group (or, in other words, we cannot define the distance form in the space of the principal fibre bundle, where fibres are given by copies of the group itself). Indeed, the generators of the Lie algebra in 4-dimensions are transformed under the $(1,0) \oplus (0,1)$ representation while the displacement one-forms $\theta^a$ are transformed under the vector $(1/2, 1/2)$ representation. Hence, the direct product of generators and the displacement forms decomposes onto a sum of some spinor representations and does not result in a trivial scalar representation which can be attributed to scalar classical distance.

**V. The variation principle for geodesics.**

The simplest Lagrangian of a classical particle that leads to the motion equations (8) is

$$S = -\int\left[\langle\phi|\hat{\theta}|\phi\rangle - d\lambda\left(\langle\phi|\phi\rangle - 1\right)\right], \qquad (23)$$

where $\lambda(\tau)$ is the Lagrange multiplier, $\langle\phi|$ is the contragradient counterpart [4] to $|\phi\rangle$, with an additional constraint that the particle state $|\phi\rangle$ is transported parallel along geodesics:

$\nabla|\phi\rangle = 0$. The particle co-ordinates are $(x, |\phi\rangle)$, and the integration is performed along the particle path $x(\tau)$ connecting the initial and final points of motion. The variation of (23) with respect to $|\phi(\tau)\rangle$, $\lambda(\tau)$ and $x(\tau)$ gives

$$\hat{\theta}(\dot{x})|\phi\rangle = \dot{\lambda}|\phi\rangle$$
$$\langle\phi|\phi\rangle = 1 \qquad (24)$$
$$\frac{d}{d\tau}\langle\phi|\hat{\theta}_i|\phi\rangle = \langle\phi|\frac{\partial\hat{\theta}}{\partial x^i}|\phi\rangle$$

The first equation of (24) with the condition $\nabla|\phi\rangle = 0$ is exactly (8), the second fixes the norm of the vector $|\phi\rangle$, the last one can be written as $\langle\phi|\nabla_{\dot{x}}\hat{\theta}_i|\phi\rangle = 0$ and comes from the fact that $\hat{\theta}$ is torsion free: $\nabla\hat{\theta}_i = 0$. The action (23) is extremised by eigenvectors (and give eigenvalues) of the distance form $\hat{\theta}(dx)$. As was mentioned above, the canonical distance form $\hat{\theta}$ has eigenvalues $d\lambda = \pm ds$, where $ds = \sqrt{\eta_{ab}\theta^a\theta^b} = \sqrt{g_{ik}dx^i dx^k}$ is the metric distance. This implies that (23) has extremum when the metric distance along the curve $x(\tau)$ has extremum, which is another definition of geodesics. This helps us to clarify why two different definitions of geodesics (as being a parallel transport of a representation vector and a minimum distance curve) coincide. Indeed, it has been proven in the section II that the distance form $\hat{\theta}$ is transported parallel along geodesics, see (9). Hence, the parallel transport of a representation vector $|\phi_0\rangle$ that extremise the distance (being eigenvalue of the canonical distance form) at the beginning of the curve will extremise the distance at any other point of the geodetic curve. So, the action (23) connects two definitions of geodesics together – it generates the parallel transport of the representation vector (as in (8)) *and* the parallel transport of the canonical distance form, which automatically extremises the metric distance along the geodetic curve.

The action (23) has a defect: it requires an additional constraint of the parallel transfer of a particle state $\nabla|\phi\rangle = 0$. Also, the action (23) contains no contribution from Yang-Mills fields. Even in the presence of non-trivial Yang-Mills potentials (which affect the covariant derivative $\nabla$), the particle of (23) will follows geodesics. A more reasonable action for the classical particle ($x(\tau), |\phi\rangle$), whose representation vector $|\phi\rangle$ is transformed by the Lorentz (orthogonal) group and a Yang-Mills group, is

$$S = -\int \left[ \langle\phi|\hat{\theta}|\phi\rangle + \frac{i}{2}\chi(\tau)\left(\langle\phi|\nabla\phi\rangle - \langle\nabla\phi|\phi\rangle\right) - d\lambda\left(\langle\phi|\phi\rangle - 1\right) \right], \tag{25}$$

where $\chi(\tau)$ and $d\lambda(\tau)$ are the Lagrange multipliers, the covariant derivative is $\nabla = d + \hat{\omega}^t = d + \hat{\omega} + \hat{\omega}^{YM}$ and once again the integration is performed along the particle path $x(\tau)$ connecting the initial and final points of motion. Here $\hat{\omega}^{YM}$ is the connection form of a local Yang-Mills group commuting with the orthogonal group. The variation of (25) with respect to $|\phi(\tau)\rangle$, $x(\tau)$, $\lambda(\tau)$ and $\chi(\tau)$ yields

$$\begin{aligned} & i\chi\nabla_{\dot{x}}|\phi\rangle + \hat{\theta}(\dot{x})|\phi\rangle = \dot{\lambda}|\phi\rangle \\ & \frac{d}{d\tau}\langle\phi|\hat{\theta}_i + i\chi\hat{\omega}_i^t|\phi\rangle - \frac{\partial}{\partial x^i}\langle\phi|\hat{\theta} + i\chi\hat{\omega}^t|\phi\rangle = 0 \\ & \langle\phi|\phi\rangle = 1 \\ & \langle\phi|\nabla\phi\rangle = 0 \end{aligned} \tag{26}$$

The particle (26) follows geodesics in the absence of the Yang-Mills fields for spaces with the constant or vanishing curvature of the linear connection $\hat{\omega}$, which can be checked by putting $\nabla_{\dot{x}}|\phi\rangle = 0$ and noting that the motion equation reduces to $\nabla_{\dot{x}}\dot{x}^i = Const \cdot R_j^i \dot{x}^j$ (with the help of identity $\langle\phi|\nabla_{\dot{x}}\hat{\omega}_i|\phi\rangle = 0$, when $\nabla_{\dot{x}}|\phi\rangle = 0$ and $\hat{\theta}(\dot{x})|\phi\rangle = \dot{\lambda}_1|\phi\rangle$).

When Yang-Mills fields are present we can simplify the motion equation using the curve calibration in which $\chi(\tau) = Const = \chi$. Simple calculations yield

$$\frac{dp_i}{dt} = \chi(R_{ij} + F_{ij})\dot{x}^j \tag{27}$$

where $p_i = \langle\phi|\hat{\theta}_i|\phi\rangle$, $R_{ij} = i\langle\phi|\hat{\omega}_{ij}|\phi\rangle$ and $F_{ij} = i\langle\phi|\hat{\omega}_{ij}^{YM}|\phi\rangle$. Combined with equations (26) (written as $i\chi\nabla_{\dot{x}}|\phi\rangle + \hat{\theta}(\dot{x})|\phi\rangle = \dot{\lambda}|\phi\rangle$, $\langle\phi|\phi\rangle = 1$ and $\langle\phi|\nabla\phi\rangle = 0$) this important equation gives a motion of a classical particle in a curved space under the action of Yang-Mills fields. We immediately see that the particle motion is affected not only by the curvature of Yang-Mills potentials $F_{ij}$ but also by the curvature of the linear connection $R_{ij}$. (As a result, a set of accelerated frames in which a particle moves along geodesic lines exists only in spaces of a constant or vanishing curvature of the linear connection. A spinor particle in the absence of Yang-Mills fields moves along the lines of $\ddot{x}^i + \Gamma^i_{jk}\dot{x}^j\dot{x}^k = \chi R_j^i \dot{x}^j$ which coincide with metric geodesics only when $R_j^i = Const$.) For spaces based on the Dirac bi-spinor representation $|\psi\rangle$ and the Abelian group of phase U(1) eq. (27) describes motion of a classical particle with spin moving in fixed external potentials. If spin is constant, the action (25) can be reduced to the action of the classical electrodynamics $S = -ds - i\chi A_\mu dx^\mu$. Here we used the antihermitian connections $A_\mu$ of mathematical literature with $A_\mu = -igA_\mu^{phys}/\chi$, where $g = e/(mc^2)$.

The action (25) is based on variables $(x, |\phi\rangle)$ which are analogous to those of a supersymmetric particle, with $x$ playing a role of "bosonic" co-ordinates and $|\phi\rangle$ - supersymmetric "fermionic" co-ordinates. It is surprising that despite this analogy the action for a supersymmetric particle is completely different from (25) and can be written in the simplest case as $S = \int g_{\mu\nu}(\dot{x}^\mu - i\langle\phi|\gamma^\mu|\dot{\phi}\rangle)(\dot{x}^\mu - i\langle\phi|\gamma^\mu|\dot{\phi}\rangle)d\tau$, see [8]. The difference is explained by different nature of spaces produced by $(x, |\phi\rangle)$ in a supersymmetric theory and here: the presence of supersymmetric transformations connecting "bosonic" co-ordinates $x$ with "fermionic" co-ordinates $|\phi\rangle$) does not allow one to consider the supersymmetric space $(x, |\phi\rangle)$ as a fibre bundle contrary to our case.

The actions (23), (25) illustrate the fact that the velocity of the Lagrange formalism can be replaced by a (spinor) representation vector as a simple alternative to momentum of the Hamilton's formalism. W. Hamilton was the first to replace a non-local velocity which enter the Lagrangian $L$ by a local momentum $p$. He also treated a momentum as an independent variable of the theory (in contrast to Lagrange velocities which are just derivative of particle position with time) because momentum definition $p = \frac{\partial L}{\partial \dot{x}}$ extremises $H$ function of the Legendre transformation $H = p\dot{x} - L$, see discussion in Ref. 10. In our case the representation vector $|\phi\rangle$ is an eigenvalue of the corresponding operator and hence also extremises the action. This allows us to treat $|\phi\rangle$ as independent variables.

It is worth mentioning that the Lagrange multiplier $\chi$ of the classical action (25) (which can be made constant by changing the curve calibration, see above) scales the "dynamical" part of the action and plays the role of the Plank constant of quantum theory [8]. There are other parallels between the classical dynamics in $(x(\tau), |\phi\rangle)$ and quantum mechanics. For example, the same direction of motion may correspond to different states of the moving particle (of different spin directions). A linear transformation of particle states $|\phi\rangle$ at some space-time points (achieved with the help of a "classical apparatus") may split particle motion. As a result, particle interference may be possible. It remains to be seen if such a classical theory will be consistent with quantum theory, for analogous attempts see [11].

**VI. The canonical distance form for spaces with the Yang-Mills group of the Standard Model.**

In case of the four-dimensional Lorentz space-time (which is an area of low energy particles and fields) the distance form is given by $\hat{\theta} = \hat{\gamma}_a \theta^a$ (or (20)) and the lowest bi-spinor representation consists of the direct sum of the left and right components. Mathematically, geodesics equation (8) represents the same geodesics as the original geodesics equation (1) provided the Yang-Mills group of the local symmetry acts identically on the left and right components. However, due to overwhelming success of the Weinberg-Glashow-Salam theory of electroweak interactions, we believe that the local symmetry group acts differently on left and right components of the orthogonal group. There exist very strong indications that the left components transform as SU(2) doublets ($L^\alpha$, $\alpha$=1,2) while the right components are SU(2) singlets ($R$). For example, the Standard Model uses $G$=SU(3)×SU(2)$_L$×U(1)$_Y$ as the local Yang-Mills symmetry group. Many other theories of grand unification also suppose different actions of SU(2) group on left and right particles. In such cases the distance form $\hat{\theta} = \hat{\theta}_{RL} \oplus \hat{\theta}_{LR}$ will connect together spaces of different irreducible representations of SU(2) group ($L^\alpha$ is a SU(2) doublet and $R$ is a SU(2) singlet) which is forbidden by the Schur's lemma [12].

It implies that the canonical distance form should be modified for SU(2)$_L$×U(1)$_Y$ particles. To remedy the situation and write the distance form for particles of representations $\begin{pmatrix} L^\alpha \\ R \end{pmatrix}$ by simplest means one needs to introduce an additional field $\varphi^\alpha$ which transforms as SU(2) doublet and glues spaces of left and right components of orthogonal representations. Then, the distance form is

$$\hat{\theta} = \hat{\theta}_{RL}\varphi^{\alpha\dagger} \oplus \varphi^{\alpha}\hat{\theta}_{LR} = \begin{pmatrix} 0 & \varphi^{\alpha}\hat{\theta}_{LR} \\ \hat{\theta}_{RL}\varphi^{\alpha\dagger} & 0 \end{pmatrix}. \qquad (28)$$

Eigenvalues of the distance form (28) can be found as

$$\hat{\theta}\begin{pmatrix} L^{\alpha} \\ R \end{pmatrix} = d\lambda \begin{pmatrix} L^{\alpha} \\ R \end{pmatrix}, \qquad (29)$$

which yields the following length element $d\lambda$:

$$d\lambda^{2} = \left|\varphi^{\alpha}\right|^{2}\eta_{ab}\theta^{a}\theta^{b} = \left|\varphi^{\alpha}\right|^{2}g_{ik}dx^{i}dx^{k}. \qquad (30)$$

The distance one-form (28) and the distance (30) generated by it are the most important suggestions of this paper. Mathematically, they are conclusions of the geodesics description in terms of particle representations and an assumption that the left and right components of the Lorentz group transform differently under the action of a local group of Yang-Mills symmetry.

The field $\varphi^{\alpha}$ obviously scales measured distance. The idea to introduce a scaling factor into the length interval is not new and was proposed some years ago by Herman Weyl in a brilliant conjecture [13] later transformed into the modern gauge theories. The original Weyl's theory has been modified in different conformal theories of gravity where dilaton fields have been introduced in order to compensate the action of local conformal symmetry group on the length element. In our case the Weyl's operator-valued one-form $\hat{\Omega}$ associated with the scaling field $\varphi^{\alpha}$ is closed. The length element (30) will not change every time the system completes a closed path, which removes the main problem of the original Weyl's theory. Field $\varphi^{\alpha}$ will be referred to as the Weyl field. It should be noted that the displacement one-forms $\theta^{a}$ are orthogonal but are not orthonormal in the presence of the Weyl field $\varphi^{\alpha}$ (which is compensated by the Weyl's form $\hat{\Omega}$ in the connection). We can introduce orthonormal displacement one-forms $\tilde{\theta}^{a} = \theta^{a}/\left|\varphi^{\alpha}\right|$; however, they will not be independent variables since they depend on the magnitude of the field $\varphi^{\alpha}$.

Writing the action (25) as

$$S = \int\left[\langle\phi|\hat{\theta}|\phi\rangle + \frac{i}{2}\chi(\langle\phi|\nabla\phi\rangle - \langle\nabla\phi|\phi\rangle) - d\lambda(\langle\phi|\phi\rangle - 1)\right], \qquad (31)$$

where $|\phi\rangle = \begin{pmatrix} L^\alpha \\ R \end{pmatrix}$, $\hat{\theta} = \begin{pmatrix} 0 & \varphi^\alpha \hat{\theta}_{LR} \\ \hat{\theta}_{RL} \varphi^{\alpha\dagger} & 0 \end{pmatrix}$, $\nabla = d + \begin{pmatrix} \hat{\omega}_{LL} & 0 \\ 0 & \hat{\omega}_{RR} \end{pmatrix} + \hat{\Omega} + \begin{pmatrix} \hat{\omega}_{LL}^{YM} & 0 \\ 0 & \hat{\omega}_{RR}^{YM} \end{pmatrix}$ ($\hat{\omega}_{LL}^{YM}$, $\hat{\omega}_{RR}^{YM}$ are the Yang-Mills connections for the left and right spinor components, $\hat{\Omega}$ is the Weyl form), we obtain the action and motion equations for a classical electron in fixed potentials of electro-weak interactions. (The contragradient counterpart $\langle\phi|$ includes a constant SU(2) vector $q^\alpha$ in order to produce a scalar $\langle\phi|\phi\rangle$.)

The distance form (28) (and the length element (30)) assigns a physical property (a value of the field $\varphi^\alpha$) to every point of space-time and support David Hume's philosophy of space-time measuring (Hume demanded a physical property to distinguish space points and used to colourise points of space). The scaling field $\varphi^\alpha$ discerns space-time points and, hence, may replace Hume's colours. The distance form (28) suggests that flat space-time (free of any matter) is an abstract mathematical notion since distance itself requires the presence of the Weyl field. Also, any form of Weyl field dynamics would provide a reason for time irreversibility in classical physics. The distance form (28) and the length element (30) might be useful in tackling some problems of classical physics such as particle existence and energy divergences of classical charged particles which will be discussed elsewhere.

## VII. Conclusion.

We study geodesics and a distance in classical physics enriched with additional co-ordinates describing particle interactions. We show that metrical relations in space-time can be determined by a canonical distance one-form acting on particle states in an associated fibre bundle. The classical dynamics based on the modified action principle describes motion of particles generated by Lorentz representations in fixed potentials of the linear connection and the Yang-Mills fields. Using concepts of the standard model of electroweak interactions we suggest that the distance form of our space-time should be scaled with a SU(2) Weyl field.

## **Appendix. A canonical distance form in a space of a Lie group.**

Eli Cartan has shown how vectors can be defined on the manifold of a Lie group ($G \times G$ principal fibre bundle) [13]. The canonical form that corresponds to these vectors is the left-invariant canonical form of the Lie group $\hat{\theta} = g^{-1}dg = \hat{X}_i \theta^i$, where $\theta^i$ are Cartan's forms and $\hat{X}_i$ are the basis of the left-invariant fields. The canonical form satisfies the Cartan-Maurer equations $d\hat{\theta} + (1/2)[\hat{\theta},\hat{\theta}] = 0$ written as

$$d\theta^i + (1/2)c^i{}_{jk}\theta^j \wedge \theta^k = 0, \qquad (32)$$

where $c^i{}_{jk}$ are the structural constant of the group. Combined with the Killing form on the Lie algebra, the canonical form $\hat{\theta}$ generates the distance on the manifold of a Lie group by $ds^2 = -\text{Tr}(ad\hat{\theta} \cdot ad\hat{\theta})$, where $ad\hat{\theta}$ is regarded as an operator on the Lie algebra acting by the adjoint representation. The torsion-free connection consistent with this metric is given by $\omega^i{}_k = (1/2)c^i{}_{jk}\theta^j$ and the torsion-free condition for this connection follows from with the Cartan-Maurer equations (32). All these facts are well known [13].

We suggest that the canonical left-invariant form of the Lie group can be regarded as the canonical distance form in accordance with the discussion above: the maximal eigenvalue of this form defines bi-invariant distance consistent with the Killing metric. Indeed, the canonical left-invariant form $\hat{\theta} = \hat{T}^{-1}(g)d\hat{T}(g)$ defines an operator in the representation space for any representation. The maximal eigenvalue of this operator is left and right invariant under the group action. The left invariance follows from the left invariance of the canonical form. The right invariance comes from the fact that the canonical form transforms by the adjoint transformation under right action of the group $R_{g'}\hat{\theta} = T^{-1}(g')\hat{\theta}T(g')$ and sets of eigenvalues of adjoint operators coincide (any eigenvalue $\lambda$ with an eigenvector $|\xi\rangle$ of the operator $\hat{\theta}$ will be an eigenvalue of the operator $R_{g'}\hat{\theta}$ with the eigenvector $T^{-1}(g')|\xi\rangle$). The maximal eigenvalue of the operator induces a distance in the operator space satisfying the triangular inequality: $\lambda_{\max}(A+B) \leq \lambda_{\max}(A) + \lambda_{\max}(B)$. (This follows from the set of inequalities:

$$\begin{aligned}\lambda_{\max}(A+B) &= \langle \xi_{\max}(A+B)|A+B|\xi_{\max}(A+B)\rangle = \\ &\langle \xi_{\max}(A+B)|A|\xi_{\max}(A+B)\rangle + \langle \xi_{\max}(A+B)|B|\xi_{\max}(A+B)\rangle \leq \\ &\langle \xi_{\max}(A)|A|\xi_{\max}(A)\rangle + \langle \xi_{\max}(B)|B|\xi_{\max}(B)\rangle = \lambda_{\max}(A) + \lambda_{\max}(B)\end{aligned}$$

where $|\xi_{max}(A+B)\rangle$, $|\xi_{max}(A)\rangle$, $|\xi_{max}(B)\rangle$ are normalised eigenvectors of the operators $A+B$, $A$, and $B$ respectively.) Thus, the eigenvalues of the canonical form of a Lie group define bi-invariant distances in the space of a Lie group provided they are not zero.

The distance generated by the canonical left-invariant form is equivalent to the distance generated by the Killing form, which follows from the uniqueness of the bi-invariant metric. Another way to see this fact is to notice that the operator space in addition to Killing form (which is essentially a Frobenius norm of operators of the adjoint representation) possesses a set of other equivalent norms. One example of the norm equivalent to the Killing metric is the numerical radius of the operator: $\max_{\langle\xi|\xi\rangle=1}|\langle\xi|A|\xi\rangle|$. For the normal matrices (which are $A^t A = AA^t$, where superscript $t$ means transposition) the spectral radius of the operator $spr(A) = \max(|\lambda||\lambda$ is an eigenvalue of $A)$ also defines a norm equivalent to the Killing metric and to the numerical norm. It means that the eigenvalues of the canonical form for many interesting groups indeed provide bi-invariant distance on the Lie group space equivalent to the traditional Killing metric.

The most attractive feature of this construction is the fact that the canonical form (which serves as the distance form) is already in the Lie algebra of the group and hence exists in the principal fibre bundle. From (32) we see that the operator-valued connection form for the metric on the Lie group can be written as $\hat{\omega} = (1/2)\hat{\theta}$. This implies that the geodesic equations (8) for the manifold of the Lie group are

$$\begin{cases} d|\xi\rangle/d\tau + (1/2)\hat{\theta}(\dot{x})|\xi\rangle = 0 \\ \hat{\theta}(\dot{x})|\xi\rangle = \lambda_{max}|\xi\rangle \end{cases}. \qquad (33)$$

The solutions of (33) are one-parameter geodesic curves [14]. This example suggests that the natural way to pack the canonical distance form into the Lie algebra is to allow the group action not only in the fibres of the corresponding fibre bundle but also on the base as well.